\RequirePackage{commath, amsmath, amssymb, amsfonts}
\documentclass[a4paper,fleqn]{cas-sc} 

\usepackage{xcolor}
\usepackage{longtable}
\usepackage{afterpage}
\usepackage{physics}
\usepackage{graphicx}
\graphicspath{{fig/}}
\usepackage[normalem]{ulem}
\usepackage{booktabs}
\usepackage{hyperref}
\usepackage{supertabular}

\definecolor{mngclr}{rgb}{0.72, 0.53, 0.04}
\definecolor{remove}{rgb}{0.5, 0.0, 0.0}

\usepackage[numbers]{natbib}
\newcommand{\obj}[1]{\texttt{#1}}

\begin{document}

\title{Atomic Simulation Recipes -- a Python framework and library for automated workflows}
\date{January 2021}

\shortauthors{Morten Gjerding et~al.}
\author{Morten Gjerding}\fnmark[1]
\author{Thorbjørn Skovhus}\fnmark[1]
\author{Asbjørn Rasmussen}\fnmark[1]
\author{Fabian Bertoldo}\fnmark[1]
\author{Ask Hjorth Larsen}\fnmark[1]
\author{Jens Jørgen Mortensen}\fnmark[1]
\author{Kristian Sommer Thygesen}\fnmark[1]
\address{$^1$Computational Atomic-scale Materials Design (CAMD), Department of Physics, Technical University of Denmark, 2800 Kgs. Lyngby Denmark}

\fntext[1]{email: mortengjerding@gmail.com}

\vspace{10pt}

\shorttitle{Atomic Simulation Recipes --- a Python framework and library for automated workflows}


\begin{abstract}
The Atomic Simulation Recipes (ASR) is an open source Python framework for working with atomistic materials simulations in an efficient and sustainable way that is ideally suited for high-throughput projects. Central to ASR is the concept of a Recipe: a high-level Python script that performs a well defined simulation task robustly and accurately while keeping track of the data provenance. The ASR leverages the functionality of the Atomic Simulation Environment (ASE) to interface with external simulation codes and attain a high abstraction level. We provide a library of Recipes for common simulation tasks employing density functional theory and many-body perturbation schemes. These Recipes utilize the GPAW electronic structure code, but may be adapted to other simulation codes with an ASE interface. Being independent objects with automatic data provenance control, Recipes can be freely combined through Python scripting giving maximal freedom for users to build advanced workflows. ASR also implements a command line interface that can be used to run Recipes and inspect results. The ASR Migration module helps users maintain their data while the Database and App modules makes it possible to create local databases and present them as customized web pages.  
\end{abstract}


\maketitle
\newpage

\section{Introduction}
As computing power continues to increase and the era of exascale approaches, the development of software solutions capable of exploiting the immense computational resources becomes a key challenge for the scientific community. In the field of materials science, ab initio electronic structure (aiES) calculations are increasingly being conducted in a high-throughput fashion to screen thousands of materials for various applications\cite{greeley2006computational,madsen2006automated,curtarolo2013high,kirklin2013high,ornso2013computational,zhang2019computational,chen2016understanding,hachmann2011harvard,bhattacharya2015high,castelli2012computational,hautier2013identification,yu2012identification,kuhar2018high,aykol2016high,mounet2018two,chen2015design} and to generate large reference data sets for training machine learning algorithms to predict fundamental materials properties\cite{rupp2012fast,lee2016prediction,xie2018crystal,ghiringhelli2015big,jorgensen2019materials,ghosh2019deep} or design interatomic potentials\cite{deringer2017machine,lorenz2004representing,behler2007generalized,artrith2016implementation}. The results from such aiES high-throughput calculations are often stored in open databases allowing the data to be efficiently shared and deployed beyond the original purpose\cite{thygesen2016making,saal2013materials,jain2013commentary,curtarolo2012aflow,draxl2019nomad,c2db,borysov2017organic,winther2019catalysis,talirz2020materials,armiento2020database,himanen2019data}. 

While a few thousands of calculations can be managed manually, a paradigm in which data drives scientific discovery calls for dedicated workflow solutions that automatically submit and retrieve the calculations, store the results in organized data structures, and keep track of the origin, history and dependencies of all data, i.e.\ the data provenance. Ideally, the workflow should also attach explanatory descriptions to the data that allows them to be easily accessed, understood, and deployed -- also by users with limited domain knowledge. 

Materials scientists from the aiES community are employing a large and heterogeneous set of simulation codes based mainly on density functional theory (DFT)\cite{kohn1965self}. These codes differ substantially in the way they implement and solve the fundamental physical equations. This is due to the fact that different types of problems require different numerical approaches, e.g.\ high accuracy vs.\ large system sizes, periodic vs.\ finite vs.\ open boundary conditions, or ground state vs.\ excited state properties. In principle, the large pool of available aiES codes provides users with a great deal of flexibility and freedom to pick the code that best suits the problem at hand. In practice, however, the varying numerical implementations and the diverse and often rudimentary user interfaces make it challenging for users to switch between the different aiES codes leading to a significant ``code barrier''. 

To some extent, a similar situation exists with respect to materials properties. Although aiES codes provide access to a rich variety of physical and chemical properties, individual researchers often focus on properties within a specific scientific domain. While this may be sufficient in many cases, several important contemporary problems addressed by the aiES community are multi-physical in nature and require properties and insights from several domains. For example, evaluating the potential of a material as a photocatalyst involves an assessment of solar light absorption, charge transport, and chemical reactions at a solid--liquid interface. Calculating new types of properties for the first time is often a time-consuming process involving trial and error and the acquisition of technical, implementation-specific knowledge of no direct benefit for the user or the overall project aim. This situation may result in a ``property barrier'' that hampers researchers' exploitation of the full capacity of aiES codes.

In this paper, we introduce The Atomic Simulation Recipes (ASR) -- a highly flexible Python framework for developing and working with computational materials workflows. The ASR reduces code and property barriers and makes it easy to perform high-throughput computations with advanced workflows while adhering to the FAIR Data Principles\cite{wilkinson2016fair}. There are already some workflow solutions available in the field, some of the most prominent being AFlow\cite{curtarolo2012aflow}, Fireworks\cite{jain2015fireworks}, AiiDA\cite{pizzi2016aiida}, and Atomate\cite{mathew2017atomate}. However, these are either designed for one specific simulation code and/or constitute rather colossal integrated entities, the complexity of which could represent an entry barrier to some users. The ASR differs from the existing solutions in several important ways, and we expect it to appeal to a large crowd of computational researchers, e.g.\ those with Python experience who like to develop their own personalized (workflow) scripts and databases, less experienced users who prefer plug-and-play solutions, and those who wish to apply non-standard methodologies, e.g.\ compute GW band structures or Raman spectra, but feel they lack the expertise required for using standard low-level codes.

The basic philosophy of ASR is to prioritize usability and simplicity over system perfection. More specifically, ASR is characterized by the following qualities:
\begin{itemize}
    \item \textbf{Flexibility:} The Python scripting interface and high degree of modularity provide users with almost unlimited freedom for developing and deploying workflows. 
    \item \textbf{Modularity:} The key components of ASR, namely the workflow development framework (ASR core), the Database and App modules, the task scheduler (MyQueue), and the simulation codes, are separate independent entities. Moreover, the Recipe library concept supports modular workflow designs and reuse of code.
    \item \textbf{Data locality:} Generated data is stored in a special folder named \texttt{.asr} where it can be accessed transparently via command line tools (similar to Git).
    \item \textbf{Compatibility:} For compatibility with external simulation codes, the ASR core is fully simulation code-independent while specific Recipe implementations communicate with simulation codes via the abstract ASE Calculator interface.    
    \item \textbf{Minimalism and pragmatism:} ASR is based on simple solutions that work efficiently in practice. This makes ASR fast to learn, easy to use, and relatively uncomplicated to adapt to future demands.   
\end{itemize}  

At the core of ASR is the concept of a \emph{Recipe}. In essence, a Recipe is a piece of code that can perform a certain simulation task (e.g.\ relax an atomic structure, calculate a Raman spectrum, or identify covalently bonded components of a material) while recording all relevant results and metadata. The use of Recipes makes it simple to run simulations from either Python or the command line. For example,
\begin{displaymath}
\verb#$ asr run "asr.bandstructure --atoms structure.json"#
\end{displaymath}
will calculate the electronic band structure of the material \texttt{structure.json}. Subsequently, the command
\begin{displaymath}
\verb#$ asr results asr.bandstructure#
\end{displaymath}
will produce a plot of the band structure. With two additional commands, the ASR results can be inspected in a web browser, see example in Fig.~\ref{fig:bandstructure}.

In practice, Recipes are implemented as Python modules building on the Atomic Simulation Environment (ASE)\cite{larsen2017atomic}. Recipes conform to certain naming and structured programming conventions, making them largely self-documenting and easy to read. To keep track of data provenance, Recipes utilize a caching mechanism that automatically logs all exchange of data with the user and other Recipes in a uniquely identifiable \texttt{Record} object. Not only does this guarantee the documentation and reproducibility of the results, it also allows ASR to determine whether a given Recipe task has already been performed (such that its result can be directly loaded and returned) and to detect if a Recipe task needs to be rerun because another piece of data in its dependency chain has changed. In addition, Recipes implement presentation and explanatory descriptions of their outputs and may also define a web panel for online presentation.  

The Recipes of the current ASR library cover a variety of computational tasks and properties (see Table~\ref{tab:recipes}). Most of the 40+ available Recipes utilize DFT. However, some Recipes do not involve calls to a simulation code (e.g.\ symmetry analysis or construction of phase diagrams) while others employ beyond-DFT methodology (e.g.\ the GW method or the Bethe--Salpeter equation). These library Recipes can be used ``out of the box'' or modified to fit the user's need. New Recipes may be developed straightforwardly following the documentation and large body of available examples. Recipes can be combined into complex workflows using Python scripting for maximal flexibility and compatibility with ASE and other relevant Python libraries like PymatGen\cite{ong2013python}, Spglib\cite{togo2018texttt} and Phonopy\cite{togo2015distributions}. The Python workflows may be executed on supercomputers using the MyQueue\cite{mortensen2020myqueue} task scheduler front-end or other similar systems. 

The ASR contains a number of tools for working with the ASE database module, which makes it easy to generate and maintain local materials databases. Relying on the Recipes' web panel implementations, these databases may be straightforwardly presented in a browser allowing for easy inspection, querying, and sharing of results on a local or public network. As an example of an ASR-driven database project we refer to the Computational 2D Materials Database (C2DB)\cite{haastrup2018computational,gjerding2021recent} \footnote{http://c2db.fysik.dtu.dk}.

While the core of ASR, i.e.\ the Recipe concept and caching system, is fully simulation code-independent, most Recipe implementations of the current library contain calls to the specific aiES code GPAW\cite{enkovaara2010electronic}. We are currently working on a generalization of the ASE Calculator interface which will decouple Recipe implementations from simulation codes. In the future, many Recipes will therefore work with multiple simulation codes.

Another on-going effort is to generalize the organization of calculated results. For example results are currently presented mainly by material.  This is practical for a database which primarily associates a number of properties with each material, but not for presenting sets of results parametrized over other variables than the material.  These limitations will be removed over the next releases.

The rest of this paper is organised as follows: In Section~\ref{sec:overview} we provide a general overview of the main components of ASR. Section~\ref{sec:recipe} zooms in on the central Recipe concept and its caching system while Section~\ref{sec:recipelibrary} gives an overview of the currently available Recipes. In Section~\ref{sec:db}, the Database and App modules are described. Section~\ref{sec:c2db} gives a brief presentation of the Computational 2D Materials Database as an example of an ASR-driven high-throughput database project and provides a few concrete examples of Recipe implementations. Section \ref{sec:interface} describes the different user interfaces supported by ASR while Sections \ref{sec:data_migration} and \ref{sec:data_provenance} explain how ASR manages data migration and provenance, respectively. Sections \ref{sec:documentation} and \ref{sec:technical specifications} cover documentation and technical specifications. Finally, Section \ref{sec:outlook} summarises the paper and presents our future perspectives for ASR.

\section{Overview of ASR}\label{sec:overview}

\begin{figure*}[t]
    \centering
    \includegraphics[width=0.95\textwidth]{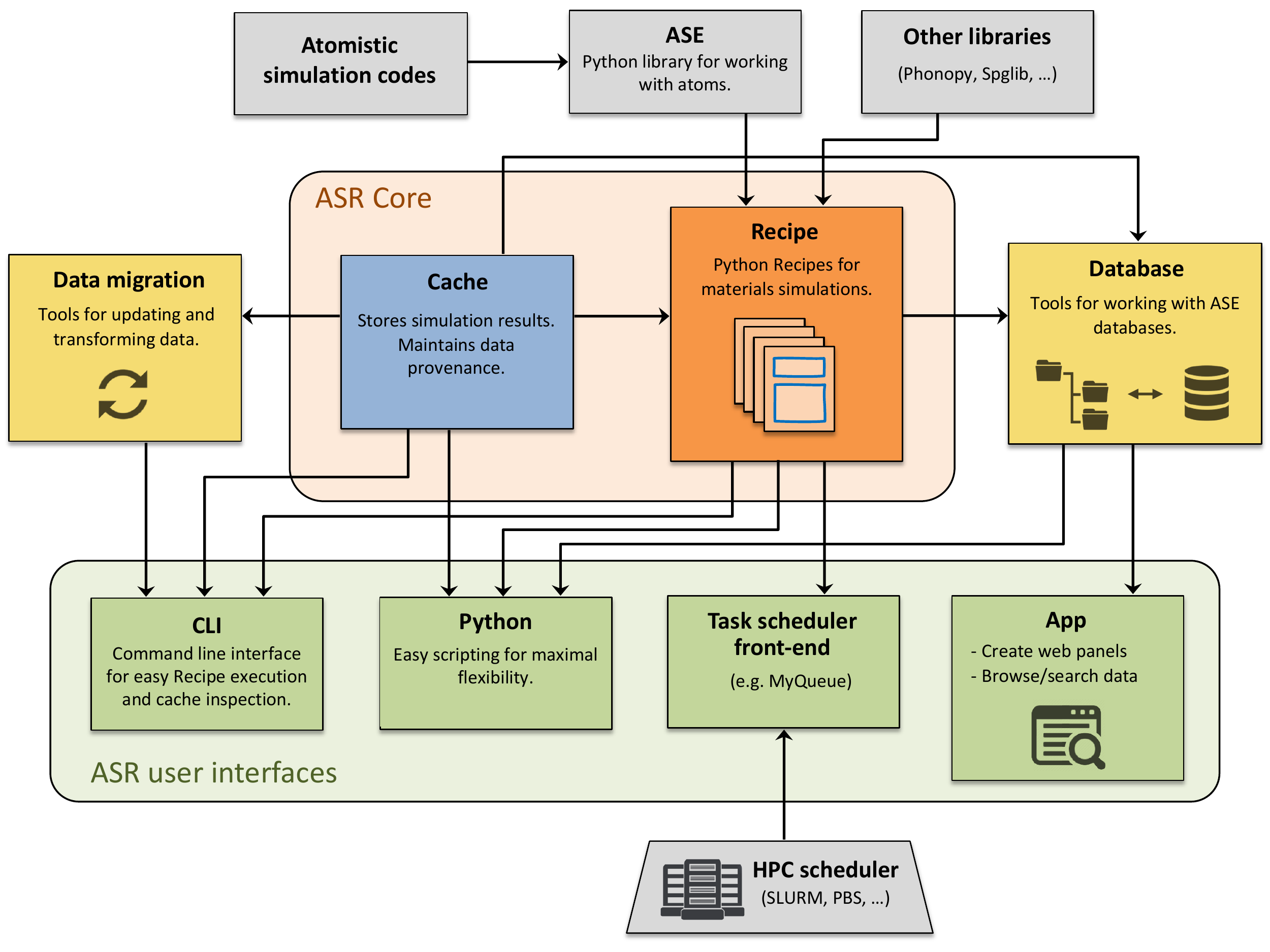}
    \caption{Schematic overview of the main modules of the ASR and their interrelations. ASR consists of a Python library of Recipes for materials simulations and a caching system for recording of results and metadata. Recipes are envisioned to communicate with simulation codes via ASE interfaces, although most current Recipe implementations contain parts that are specific to the GPAW code. An arrow from X to Y means that Y calls X. The blue frames on the Instructions Recipe box symbolise a caching layer that records all data flow to/from the Recipes.  
    }
    \label{fig:overview}
\end{figure*}

Fig.~\ref{fig:overview} shows a schematic overview of the main components of the ASR and their mutual dependencies. An arrow from X to Y indicates a direct dependence of Y on X, e.g.\ via function calls (Y calls X). The ASR modules have been divided into the ASR core modules (Cache and Recipe) and the ASR user interfaces (command-line interface, Python, Task scheduler front-end, and Apps). In addition, the ASR Database and Data migration modules contain tools for working with databases and maintaining data, respectively. 

Recipes implement specific, well defined materials simulation tasks as Python modules building on the ASE\cite{larsen2017atomic} and other Python libraries. A Recipe integrates with a Cache module that keeps track of performed tasks and manages all relevant metadata. The Cache also allows the user to inspect the data generated by a Recipe via the ASR command line interface (CLI) or using Python. Likewise, the Recipes may be executed directly from the CLI or called via Python scripts, the latter giving maximal flexibility and compatibility with existing Python libraries. For the purpose of high-throughput computations, advanced Python workflows combining several Recipes may be constructed and executed remotely using task scheduling systems like MyQueue\cite{mortensen2020myqueue}.   

The ASR Cache and Recipe modules work on a folder/file basis. This locality of data makes the ASR highly transparent for the user. The ASR Database module contains functions for converting the ASR data stored in a tree of folders into an ASE database and vice versa. The ASR App module generates web pages for online presentation, browsing and searching of the databases generated by the ASR Database module. Finally, the Data migration module provides tools for transforming data (results or metadata) to ensure backward compatibility when Recipes are updated.

\section{What is a Recipe?}\label{sec:recipe}

\begin{figure*}[t]
    \centering
    \includegraphics[width=0.9\textwidth]{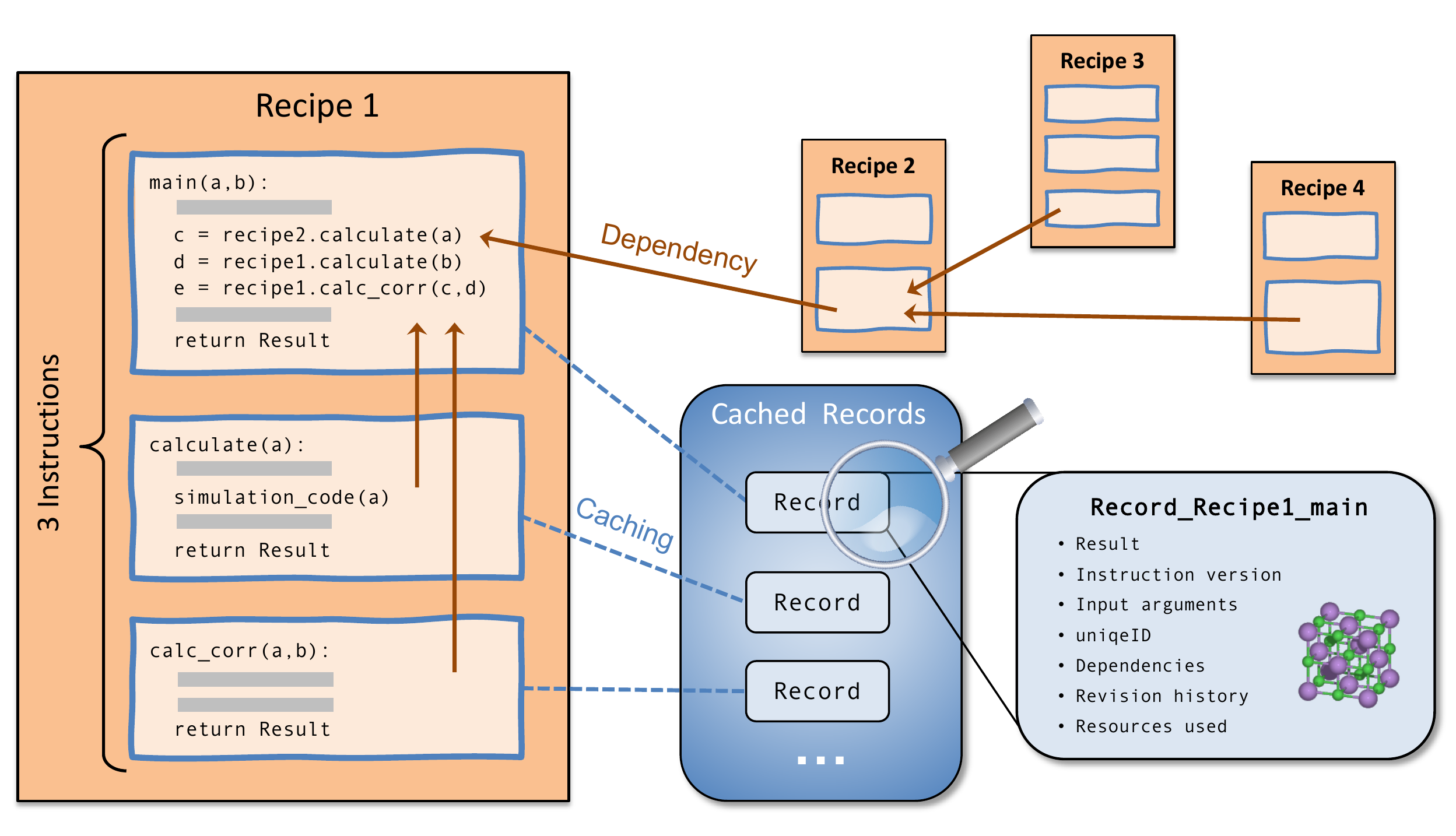}
    \caption{
   A Recipe consists of a set of Instructions (see Fig.~\ref{fig:instruction}) implementing the computational steps needed to obtain a desired result. An Instruction may call other Instructions of the same, or separate, Recipes. An Instruction always returns a \texttt{Record} holding its result, normally represented as a \texttt{Result} data structure, together with the dependencies on other Instructions and all additional metadata required to trace back and reproduce the result.  
    }
    \label{fig:recipe_structure}
\end{figure*}

\begin{figure*}[h]
    \centering
    \includegraphics[width=0.9\textwidth]{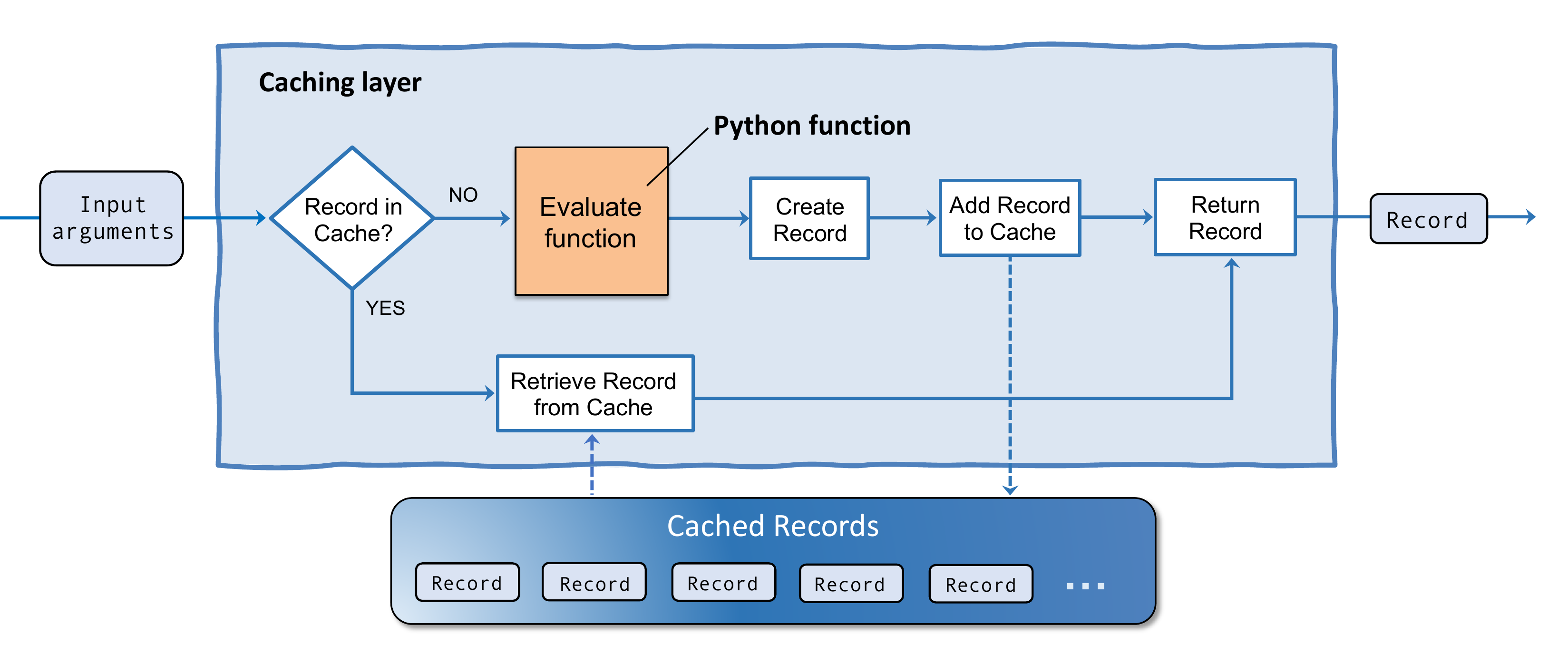}
    \caption{
    An Instruction is a Python function (orange) wrapped in a caching layer (light blue). When the function is called with a set of input arguments, the caching layer consults the cache to check if a \texttt{Record} for that exact function call already exists. In case of a cache hit, the \texttt{Record} is read and returned. In case of a cache miss, the function is evaluated and the \texttt{Record} is stored before it is returned. 
    }
    \label{fig:instruction}
\end{figure*}

A Recipe is a Python module implementing the Instructions needed to obtain a particular result, for example to relax an atomic structure, calculate an electronic band structure or a piezoelectric tensor. This section describes the structure and main components of a Recipe. A schematic overview of the Recipe concept is shown in Fig.~\ref{fig:recipe_structure}. 

\subsection{Instruction} An Instruction is to be understood as a Python function wrapped in a caching layer provided by ASR, see Fig.~\ref{fig:instruction}. Whenever an Instruction is called, the caching layer intercepts the input arguments and asks the cache whether the result of the particular Instruction call already exists (cache hit) or whether there are no matching results (cache miss). If a matching result exists (because it was calculated previously), the caching layer skips the actual evaluation of the Instruction and simply reads and returns the previously calculated result. In the case of a cache miss, the Instruction is evaluated, after which the result is intercepted by the caching layer and stored together with the relevant metadata in a \texttt{Record} object. The precise content of the \texttt{Record} object and the conditions for a cache hit/miss are described in Section \ref{sec:record}. 

One of the great benefits of this design is its simplicity. Because the Instruction/caching layer is implemented as a simple wrapper around a Python function, usage of the caching functionality requires minimal additional knowledge. In practice, this means that working with ASR and implementing new ASR Recipes becomes really simple. 

The caching system works on a per-folder basis (similar to Git): a cache is initialized by the user in a folder and any instruction evaluated within this folder or sub-folders will utilize this cache. This mimics the behaviour of the MyQueue task scheduler so as to maximize the synergy between these tools. In practice, the ``one-cache-per-folder'' system works well together with a ``one-material-per-folder'' structure. The latter is currently still a requirement for utilizing the Database functionalities described in Section~\ref{sec:db}. However, the caching system can work with several atomic structures in the same folder as the cache can distinguish ASR tasks performed on different atomic structures. 
Data written by ASR is encoded as \texttt{JSON}.

 Any Instruction can be called directly by the user (from Python or the CLI), but special importance is given to the ``main Instruction''. The main Instruction usually provides the primary interface for the user to the Recipe and returns the final result of the Recipe. Other Instructions are called by the main Instruction and evaluated as needed. These may be Instructions implemented in the Recipe itself but may also be Instructions of separate Recipes. The main Instruction takes all input arguments required by the Recipe and uses them to call other Instructions.

Having multiple Instructions in a Recipe is usually motivated by code reusability or reduction of resources. The former is relevant when another Recipe needs to perform an identical Instruction (see Section on \emph{Dependencies}). The latter is relevant when the task can be divided into Instructions with different resource requirements, in which case the separation may save computational time or resources. In particular, this is useful if a recalculation of a subset of the generated data is required. 

The \emph{input arguments} of an Instruction comprise all the information required to specify its task. When calls to an external simulation code are involved, the input arguments include a code specification, the computational parameters like $k$-point density, basis set specification, or exchange--correlation (xc) functional, as well the atomic structure. 

An Instruction carries a \emph{version number} 
to facilitate data migrations, i.e.\ transformations of the values or organisation of data produced by the Instruction.  This may be required for backward compatibility when Instructions are updated, see Section \ref{sec:data_migration}.

\subsection{Dependencies}\label{sec:dependencies}
It often happens that an Instruction can benefit from the functionality implemented by other Instructions. An example is the main Instruction of the ``band structure'' Recipe which calls an Instruction of the ``ground state'' Recipe to compute the electron density that the band structure should be based on. The caching layer logs whenever an Instruction requests data from another Instruction and uses that information to build a list of data dependencies. The data dependency list is stored in the \texttt{Record} object making it possible to trace what other pieces of data were used in the construction of the current result. 

Implementation of data-dependencies in Recipes requires no extra coding. Whenever an Instruction calls another Instruction, the caching layer will \emph{automatically} intercept the call and (1) determine if there exists a matching Record (cache hit/miss); (2) log the data dependency by registering the unique IDs and revision UIDS (see Section \ref{sec:data_migration}) of any dependent Records.

\subsection{The \texttt{Record} object}
\label{sec:record}
The \texttt{Record} object is the basic data unit of ASR. It stores the results of Instructions together with metadata documenting how the results were obtained, and is used by the cache system to identify already performed Instruction calls. The \texttt{Record} object  
contains the following information:
\begin{itemize}
    \item \texttt{Result} object (see Section \ref{sec:result})
    \item Input arguments, if relevant including
    \begin{itemize}
    \item Atomic structure
    \item Simulation code specification
    \item Computational parameters 
    \end{itemize}
    \item Instruction version (see Section \ref{sec:data_migration})
    \item External codes versions
    \item Randomly generated unique ID
    \item Dependencies (see Section \ref{sec:dependencies})
    \item Revision History (see Section \ref{sec:data_migration})
    \item Execution time and resources (number of cores)
\end{itemize}

To identify a cache hit/miss when evaluating an Instruction, the caching layer searches the cache for \texttt{Records} with matching Instruction name, version, and input arguments. A cache hit is then defined as the existence of a matching \texttt{Record}. A recursive comparison is used to
 compare input arguments with those from existing \texttt{Records} within a small
 numerical tolerance for floating point numbers.
Any later evaluations of the Instruction with identical arguments will result in a cache hit. 

\subsection{The \texttt{Result} object}\label{sec:result}
To store and document the result produced by a Recipe, ASR offers a \texttt{Result} object that wraps the actual result data (stored as a Python dictionary) in a simple data structure that also contains specification of the result data types along with short explanatory descriptions of the data. In addition, the \texttt{Result} object may implement methods to present itself in different formats, see below. Using the \texttt{Result} object is optional, but in practice all instructions that return more than a simple object or value utilizes a \texttt{Result} object for improved data documentation.

\subsection{Presentation of results}
The \texttt{Result} object may implement presentation options of the result data in various formats, for example text to terminal, figures, and web panels. The ASR Database and App modules draw on the Recipes' web panel implementation to create web pages for presenting, browsing, and distributing databases containing collected \texttt{Result} objects, see sections \ref{sec:db} and \ref{sec:c2db}. This provides an efficient way of inspecting and sharing data as it is generated, which is highly practical for projects involving multiple collaborators. 

\subsection{General principles for Recipe development}\label{sec:guideline}
To maintain and exploit the modular structure of ASR, the development of new Recipes should follow a few general design principles. First, the task performed by a Recipe should be well defined and clearly bounded to make it easy to use in different contexts. It should always be considered whether the Recipe could be split into smaller independent Recipes that could be useful individually. Additionally, it is encouraged that Recipes are designed/programmed so as to be as broadly applicable as possible, e.g.\ with respect to the type of material (structure dimensionality, chemical composition, magnetic/non-magnetic, metallic/insulating, etc.). Any information required to define the simulation task should be included in the input argument of the Recipe, i.e.\ hard coding of parameters should be avoided. This should be done to ensure a flexible use and enhance the data provenance (input arguments are stored in the \texttt{Records}). Recipes should employ conservative parameter settings as default to ensure that the results are numerically well converged independent of the application, e.g. material type. Finally, in order to keep ASR Recipes simple and easy-to-read, and in order to enhance the modularity, code-extensive functionalities should be separated out into ASE functions and called from ASR whenever it is possible and sensible, i.e.\ when the ASE function is useful in other contexts than the specific Recipe.

\section{The Recipe library}\label{sec:recipelibrary}
The ASR currently provides more than 40 complete Recipes allowing users to perform a broad range of materials simulation tasks ranging from construction and analysis of crystal structures over DFT calculations of thermodynamic, mechanical, electronic, magnetic, and optical properties to many-body methods for evaluating response functions, quasiparticle band structures, and collective excitations. A non-exhaustive list of available Recipes is provided in Table~\ref{tab:recipes}. It should be stressed that the list constitutes a snapshot of the current state of the Recipe library, which is continuously expanding. For example, we are currently developing Recipes for creating and modeling layered van der Waals structures and point defects in semiconductors. 

Most of the currently implemented Recipes rely specifically on the GPAW\cite{enkovaara2010electronic} electronic structure code. As previously mentioned, we are currently working on a generalisation of the ASE Calculator interface to make the Recipes -- or a large portion of them -- simulation code-independent. Until then, usage of ASR with other simulation codes than GPAW is possible by porting of existing Recipes or development of new ones. The amount of work involved will depend on the type of Recipe and the state of the ASE interface for the specific simulation code.   

A few specific examples of Recipe implementations are given in Section \ref{sec:c2db} where we outline the main computational steps and the final output of the \texttt{asr.bandstructure} and \texttt{asr.emasses} Recipes, respectively. 

\onecolumn
\begin{table}[h!]
    \centering
    \caption{List of Recipes currently implemented in the ASR library. Most of the Recipes depend explicitly on the GPAW electronic structure code. The Recipes are grouped under thematic headings and listed in alphabetic order.}
    \label{tab:recipes}
    \begin{supertabular}{ll}
        \toprule
        Recipe name  & Description \\ 
        \hline
        \vspace{-6pt} & \\
        \textbf{Atomic structure} & \\
        asr.database.duplicates & Remove duplicate structures from a database \\
        asr.database.rmsd & Root mean square distance between structures  \\
        asr.dimensionality & Dimensionality of covalently bonded substructures of a material \\
        asr.push & Push atoms along specific phonon mode \\
        asr.relax & Relax atomic structure \\
        asr.setup.defects & Generate native point defects \\
        asr.setup.displacements & Generate structures with a single displaced atom \\
        asr.setup.magnetize & Initialize atomic magnetic moments \\
        asr.setup.reduce & Reduce supercell to primitive cell \\
        asr.setup.symmetrize & Symmetrize an atomic structure \\
        asr.structureinfo & Extract structural information \\
        \vspace{-6pt} & \\
        \textbf{Thermodynamic properties} & \\
        asr.chc & Constrained convex hull stability analysis \\
        asr.convex\_hull & Convex hull stability analysis \\
        asr.defectformation & Formation energy of neutral point defect \\
        asr.fere & Define elemental reference energies \\
        \vspace{-6pt} & \\
        \textbf{Mechanical properties} & \\
        asr.phonopy & Phonon band structure and dynamical stability  \\
        asr.piezoelectrictensor & Piezoelectric tensor \\
        asr.stiffness & Stiffness tensor \\
        \vspace{-6pt} & \\
        \textbf{Electronic properties} & \\
        asr.bader & Bader charge analysis \\
        asr.bandstructure & Kohn-Sham band structure \\
        asr.berry & Various band topology invariants \\ 
        asr.borncharges & Born effective charge tensor \\
        asr.deformationpotentials & Deformation potentials (only for 2D) \\
        asr.dos & Density of states \\
        asr.emasses & Effective masses \\
        asr.fermisurface & Fermi surface \\
        asr.formalpolarization & Formal polarization phase \\
        asr.gs & Electronic ground state \\
        asr.gw & G$_0$W$_0$ quasiparticle band structure \\
        asr.hse & HSE06 band structure \\
        asr.pdos & Orbital projected density of states \\
        asr.projected\_bandstructure & Orbital projected Kohn--Sham band structure \\
        \vspace{-6pt} & \\
        \textbf{Magnetic properties} & \\
        asr.exchange & Magnetic exchange coupling \\
        asr.magnetic\_anisotropy & Magnetic anisotropy \\
        asr.magstate & Determine magnetic state \\
        \vspace{-6pt} & \\
        \textbf{Optical properties} & \\
        asr.bse & Optical absorption from Bethe--Salpeter Equation (BSE) \\
        asr.infraredpolarizability & Infrared polarizability (caused by vibrations) \\
        asr.plasmafrequency & Plasma frequency (from intraband transitions) \\
        asr.polarizability & Optical polarizability (caused by electrons) \\
        asr.raman & Raman spectrum (first-order) \\
        asr.shg & Second harmonics generation  \\
        asr.shift & Shift current \\
        \toprule
    \end{supertabular}
\end{table}

\section{The ASR Database and App modules}\label{sec:db}
The ASR Database and web App modules make it possible to package, inspect, share, and present ASR-driven projects easily and efficiently. The main tools and opportunities provided by these modules are described in more detail below.

\subsection{Database}
The ASR Database module can be used to collect \texttt{Record} objects from a directory tree into an ASE database. This is achieved by the command \texttt{asr database fromtree}. The procedure assumes a ``one-material-per-folder'' structure, relying on the existence of an atomic structure file in each folder to select \texttt{Records} pertaining to that atomic structure. The Database module proceeds to collect atomic structure-\texttt{Record} data sets and assign them to a particular row of an ASE database. We shall refer to such a database as an ASR database. Once an ASR database has been collected, it is possible to define key--value-pairs and relate property data to specific atomic structures.

The Database module also enables the reverse operation, that is, unpacking an existing ASR database to a directory tree containing \texttt{Record} objects. This is achieved by the command \texttt{asr database totree}. The function is useful when continuing a project, e.g.~because existing data must be updated or new
data must be added, for which the database is available but not the original directory tree. Moreover, it provides tools for merging and splitting databases.

It is possible to collect a database for any number of materials/\texttt{Record} objects -- even for a single material -- and thereby take advantage of the App tools for presenting and inspecting results in a browser with no extra efforts. However, collecting databases is obviously most powerful in cases involving many materials/properties where the database makes it possible to search and filter the data via the defined key--value-pairs. 

The easy installation of ASE through the standard PyPI Python package manager makes the ASE database format highly accessible. Furthermore, the portability of an ASE database (via several backends, e.g.\ SQLite, PostgreSQL, MariaDB and MySQL) enables easy packaging and distribution of data among different parties. 

\subsection{Web App}
The ASE provides a flexible and easily extensible database web application making it possible to present and inspect the content of an ASE database in a browser. ASR leverages this ASE functionality to customize the web application layout and provide more sophisticated features such as the automatic generation of web panels, generation of figures, and documentation of the presented data by utilizing the web panel data structures encoded in the \texttt{Result} objects. Normally a Recipe generates one web panel. However, panels gathering data from several Recipes may be created. One example of the latter is the ``Summary'' panel of the C2DB web pages discussed in the next section. In this case, a number of Recipes write data to a web panel data structure named ``Summary'' in their \texttt{Result} object. This information is stored in the database when collected. When generating the C2DB web pages from the C2DB database, the App constructs all web panels that are defined in the data pertaining to a particular material. If several Recipes have written to the same web panel, the data will be combined in an order controlled by a priority keyword written together with the web panel data. 

\subsubsection{Adding information fields}
To enhance the accessibility of the data, it is possible to add an explanatory description to specific data entries, i.e, key--value pairs and data files, of an ASR database. These descriptions will appear as text boxes when clicking a ``?''-icon placed next to the data on the web panels, see Fig.~\ref{fig:summary}. General information boxes for web panels are always generated by ASR. They contain a customised field that can be manually edited, e.g.\ providing a short explanation of the data presented in the panel and/or links to relevant literature, and an automatically generated field listing the ASR Recipes that have produced data for the web panel and the key input parameters for the calculations. An example of such an information box is shown in Fig. \ref{fig:summary}.

\subsubsection{Linking rows of databases} 
ASR provides functionality to create links between rows of the same, or different, ASR databases. This allows the developer to connect relevant materials when designing web panels such that the end user can move swiftly between them when browsing databases. For example, the \texttt{asr.convex\_hull} Recipe creates the convex hull phase diagram of a material using an ASR reference database of stable materials (originally from from the OQMD\cite{saal2013materials}), and creates a table with links to all the materials on the phase diagram. Other examples, could be to link different defective versions of the same crystalline material or different isomers of the same material/molecule. 

The links are defined in \texttt{links.json} files in the folders of the relevant materials. These files may be generated manually or automatically using the Recipe \texttt{asr.database.treelinks}. When collecting the database, ASR reads the \texttt{links.json} file for each folder and stores the information in the \texttt{Data} dictionary of the corresponding row. The Recipe \texttt{asr.database.crosslinks} then creates links between rows of the collected database and rows of other databases that are given as input to the Recipe. When generating the web panels, ASR uses this information to generate hyperlinks in \texttt{HTML} format and present them in the web application for each material.

\section{High-throughput example: The C2DB}\label{sec:c2db}
In this section we present an example of what can be accomplished by the ASR in the realm of data intensive high-throughput applications, showcase some examples of ASR-generated web panels,  and discuss two specific Recipe implementations. 

Historically, the ASR evolved in a symbiotic relationship with the Computational 2D Materials Database (C2DB) --- an extensive database project organising various properties of more than 4000 two-dimensional (2D) materials. 
The C2DB distinguishes itself from existing computational databases of bulk\cite{saal2013materials,jain2013commentary,curtarolo2012aflow} and low-dimensional\cite{ashton2017topology,mounet2018two,zhou20192dmatpedia} materials by the large number of physical properties available. These include convex hull diagrams, stiffness tensors, phonons (at high-symmetry points), projected density of states, electronic band structures with spin--orbit effects, effective masses, band topology indices, work functions, Fermi surfaces, plasma frequencies, magnetic anisotropies, magnetic exchange couplings, Bader charges, Born charges, infrared polarisabilities, optical absorption spectra, Raman spectra, and second harmonics generation spectra. The use of beyond-DFT theories for excited state properties (GW band structures and BSE absorption for selected materials) and Berry-phase techniques for band topology and polarization quantities (spontaneous polarization, Born charges, piezoelectric tensors), are other unique features of the C2DB. 

Building the first version of C2DB without a fully functioning workflow framework was a long and painstaking endeavour, but absolutely critical for the successful development of the ASR. 
Today, the entire C2DB project can be generated by a single (MyQueue) Python workflow script comprising a sequence of ASR Recipe calls and simple Python code for controlling and directing the workflow via statements like ``\texttt{if band\_gap > 0:}''. Relying on the MyQueue task scheduler (see Section \ref{sec:myqueue}), generation of the C2DB is accomplished by the single command ``\texttt{mq workflow c2db\_workflow.py tree/*/*/*/}'', which will submit the C2DB workflow in folders matching the pattern \texttt{tree/*/*/*/}. With the current C2DB workflow, this statement will launch up to 23 unique Instructions for each of the 4047 materials amounting to a total of 59822 individual aiES calculations (some Recipes like phonon and stiffness calculations launch multiple aiES calculations). When the current workflow is run with the GPAW code, about 258 calculations are unsuccessful (most often due to convergence errors in the self-consistency DFT cycle) corresponding to a success rate of 99.5\%.

Apart from the data provenance control that ensures the documentation and reproducibility of the data, there are two aspects of the ASR that are particularly crucial for making high-throughput computations work efficiently in practice. First, the caching functionality ensures that Recipes which have already been performed are automatically skipped by ASR (unless something in the input for a Recipe has changed since it was last executed). This means that only a single workflow script needs to be maintained and submitted every time something has been changed, e.g., new materials have been added, the workflow script has been updated, it has been decided to rerun certain tasks with new parameters, or a Recipe has been modified. Such functionality is essential because running and maintaining high-throughput projects inevitably requires that subsets of calculations are repeated at different points in time. Secondly, the carefully designed and well tested Recipes minimise the number of unsuccessful calculations and the risk of human errors. 

\begin{figure*}[t]
    \centering
    \includegraphics[width=0.9\textwidth]{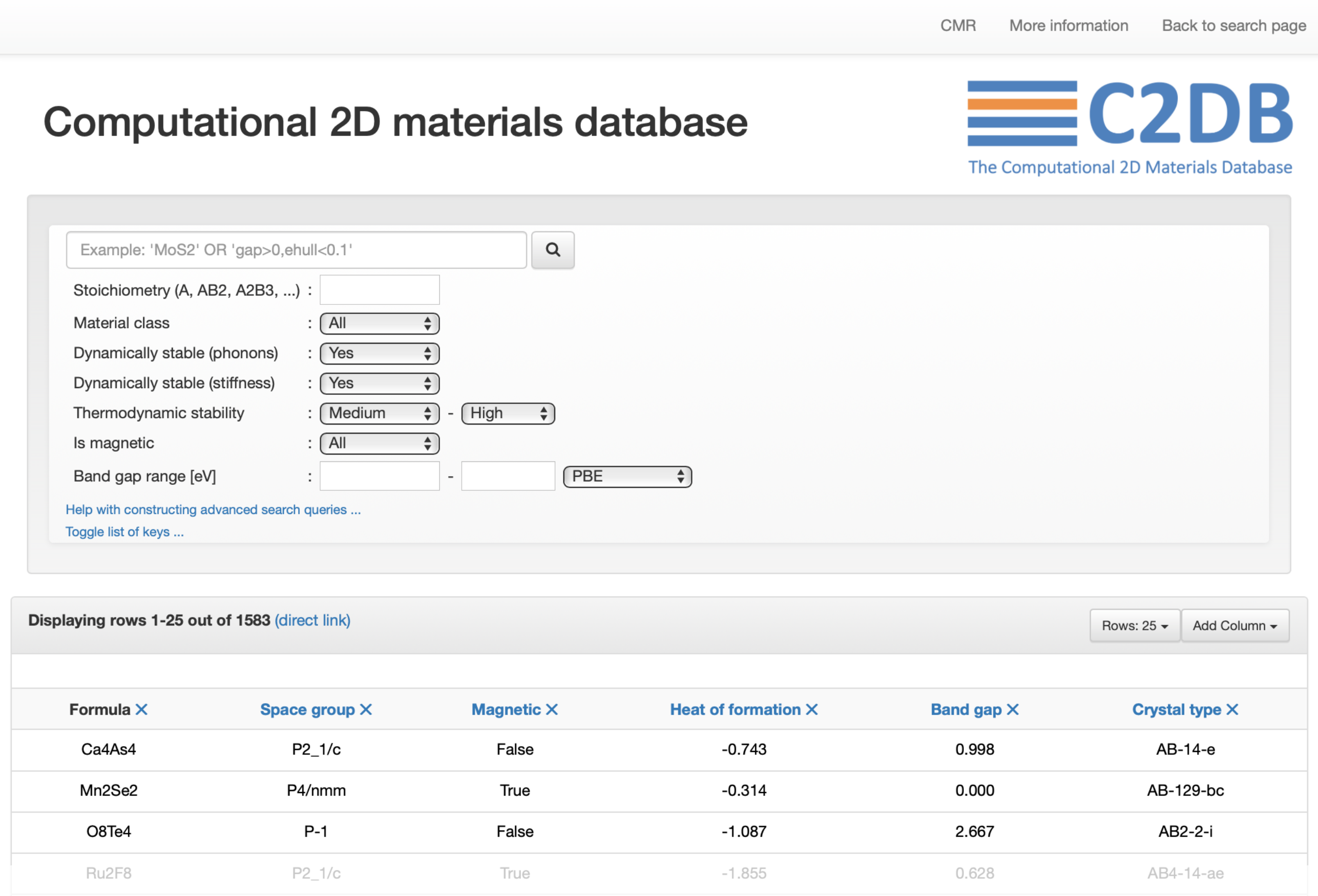}
    \caption{The search page of C2DB with the first few rows of the database shown below. The default web page generated by ASR includes only the top most search field, but the panel can be customized by additional fields and buttons for more convenient data filtering.}
    \label{fig:search}
\end{figure*}
\begin{figure*}[h!]
    \centering
    \includegraphics[width=0.78\textwidth]{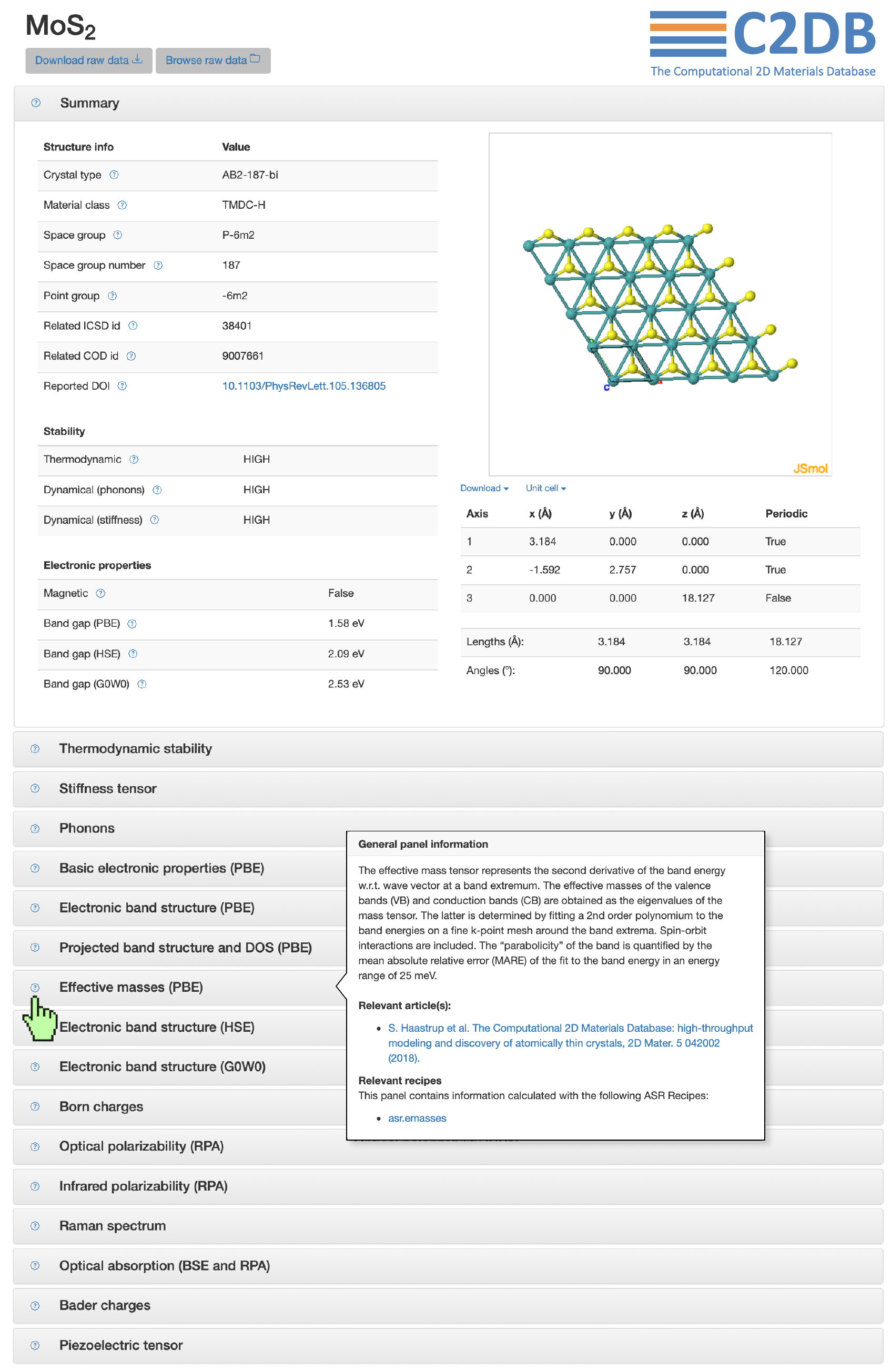}
    \caption{Screenshot of the web page for monolayer MoS$_2$ from the C2DB project (only the ``Summary'' panel is unfolded). The panel presents data from the \texttt{Result} objects generated by the following Recipes: \texttt{asr.gs}, \texttt{asr.gw}, \texttt{asr.hse}, \texttt{asr.phonons}, \texttt{asr.magstate}, \texttt{asr.stiffness}, \texttt{asr.convex\_hull}, \texttt{asr.structureinfo}.}
    \label{fig:summary}
\end{figure*}

\subsection{Recipe and web page examples}
 Below we present a few examples of output generated by the ASR-C2DB workflow (for a full impression we refer the reader to the C2DB website). 
 
 \subsubsection{Search page}
Fig.~\ref{fig:search} shows the C2DB search page, which consists of a search/filtering section followed by a list of the database rows presented by a selected number of key--value pairs. Clicking one of the highlighted key names once (twice) will sort the rows in increasing (decreasing) order of that key. Which keys should be shown by default can be customized, but the user can always add extra keys via the ``Add column'' button. By default, 
the search page generated by the ASR App module will contain only the search field in the upper section, but additional fields or buttons may be added for easy filtering according to the most relevant parameters. 

\subsubsection{``Summary'' panel}
 Fig.~\ref{fig:summary} shows the C2DB web page for monolayer MoS$_2$. All the web panels produced by the various Recipes of the workflow are seen, but only the ``Summary'' panel is unfolded. This panel is designed to provide an overview of the most basic properties of the material, and gathers data from the \texttt{Result} objects generated by the following Recipes: \texttt{asr.gs}, \texttt{asr.gw}, \texttt{asr.hse}, \texttt{asr.phonons}, \texttt{asr.magstate}, \texttt{asr.stiffness}, \texttt{asr.convex\_hull}, and \texttt{asr.structureinfo}. 
 
 Fig.~\ref{fig:summary} also shows the information box of the ``Effective masses'' web panel. It contains a short explanation of the effective mass tensor and how it is evaluated by the Recipe as well as a link to a relevant paper. The automatically generated part shows that the panel contains data generated by the \texttt{asr.emasses} Recipe.    
  The two fields at the top of the page ``Download raw data'' and ``Browse raw data'' provide access to the entire data set comprised by all \texttt{Result} objects of the specific material entry of the database.

\subsubsection{``Band structure'' Recipe}
\begin{figure*}[!h]
    \centering
    \includegraphics[width=\textwidth]{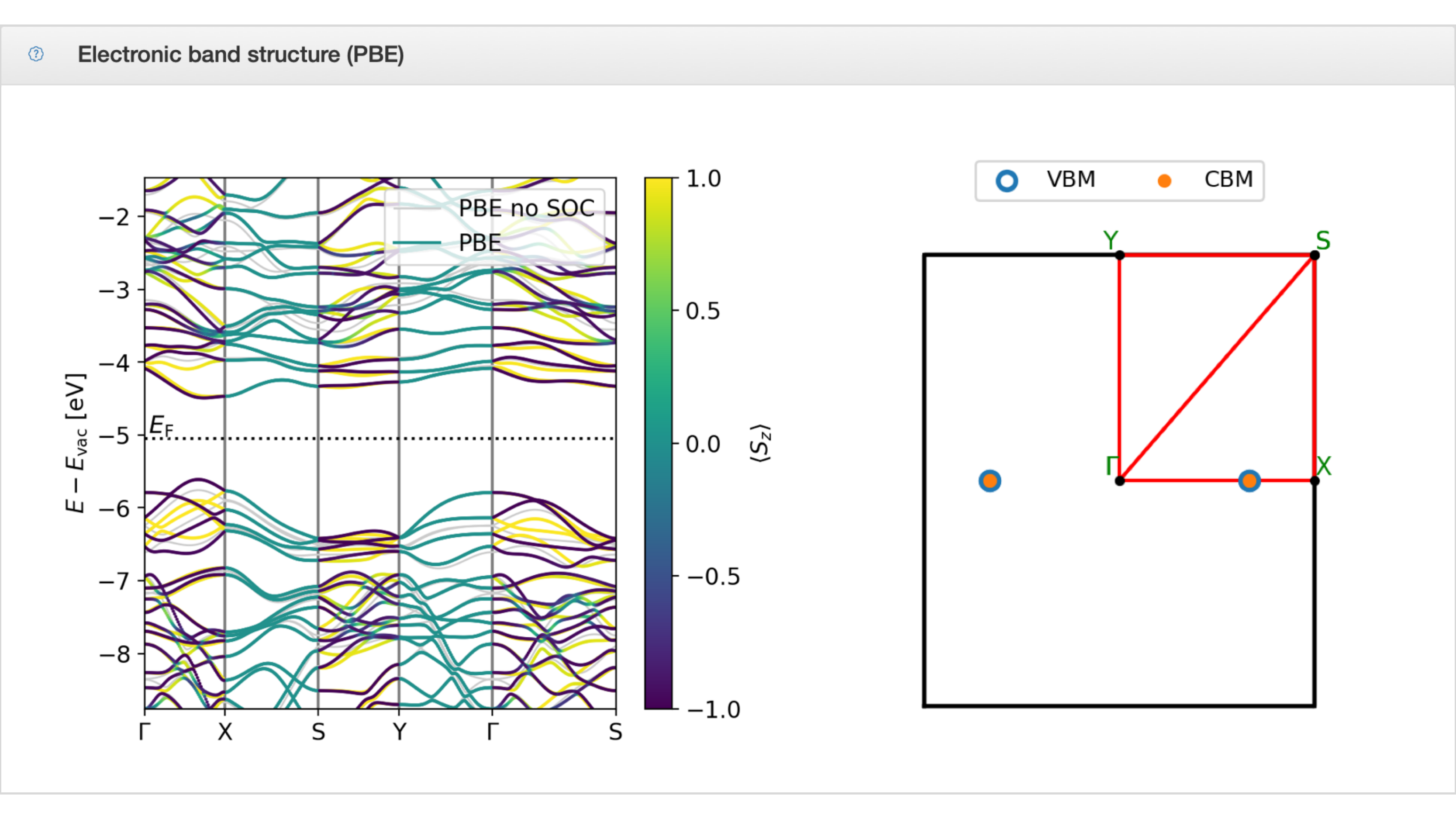}
    \caption{Screenshot of the ``Band structure'' panel for monolayer CrW$_3$S$_8$ from the C2DB project. The web panel contains data computed by the \texttt{asr.bandstructure} Recipe.}
    \label{fig:bandstructure}
\end{figure*}

As another example, Fig.~\ref{fig:bandstructure} shows the ``Electronic band structure'' panel for monolayer CrW$_3$S$_8$ as calculated and presented by the Recipe \texttt{asr.bandstructure}. The band structure is calculated with the PBE xc-functional including spin--orbit interactions. The out-of-plane spin projections of the states is shown by the color code. The main computational steps carried out by this Recipe are: 
\begin{itemize}
    \item Perform a self-consistent ground state calculation (by calling the \texttt{calculate} Instruction of the ground state Recipe \texttt{asr.gs}) to obtain a converged electron density.
    \item Determine crystal symmetries and corresponding band path (uses ASE functionalities).
    \item Calculate the Kohn--Sham eigenvalues along the band path. For magnetic materials, this step calls the Recipe \texttt{asr.magnetic\_anisotropy} to obtain the magnetic easy axis for evaluating spin projections. 
    \item Call the main Instruction of the ground state Recipe to get the Fermi level (in 3D) or the vacuum level (in <3D) for use as zero-point energy for the band structure.
\end{itemize}
In addition to these computational steps, the main Instruction of the Recipe formats two figures to present the band structure itself and the Brillouin zone with the band path and the positions of the valence band maximum (VBM) and conduction band minimum (CBM). Note that the position of the VBM and CBM, as well as a number of other properties like the band gap and band edge energies (not shown), are determined by the Recipe \texttt{asr.gs}, which is called by \texttt{asr.bandstructure}.

\subsubsection{``Effective masses'' Recipe}
\begin{figure*}[!h]
    \centering
    \includegraphics[width=\textwidth]{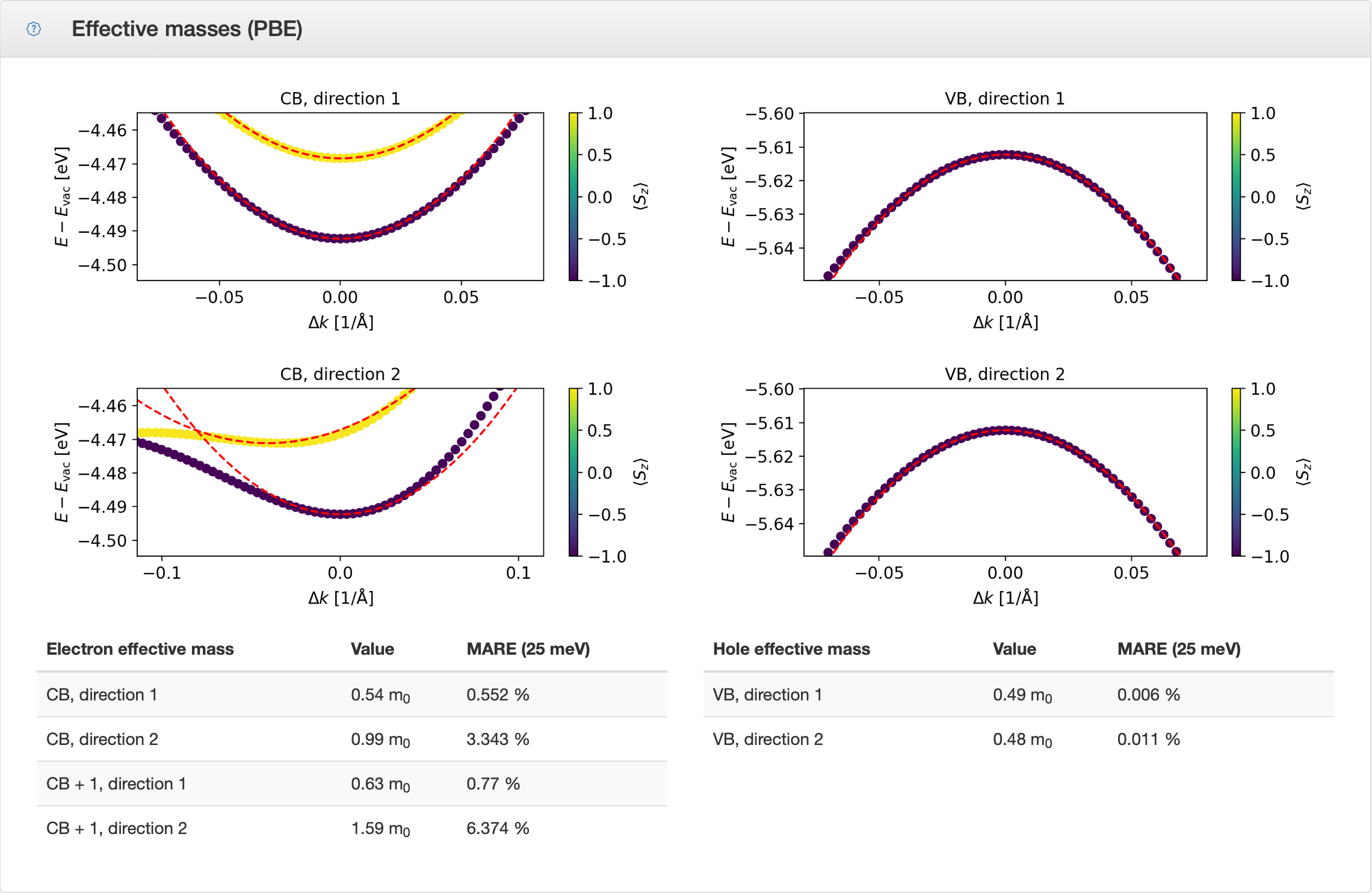}
    \caption{Screenshot of the ``Effective masses'' panel for monolayer CrW$_3$S$_8$ from the C2DB project. The panel contains data computed by the \texttt{asr.emasses} Recipe.}
    \label{fig:emasses}
\end{figure*}

Fig.~\ref{fig:emasses} shows a screenshot of the ``Effective masses'' panel for monolayer CrW$_3$S$_8$ generated by the Recipe \texttt{asr.emasses}. The effective mass tensor is calculated with the PBE xc-functional including spin--orbit interactions. The color code represents the spin projections along the $z$-axis. In addition to the effective masses themselves, the Recipe evaluates a ``band parabolicity'' parameter defined as the mean absolute relative error (MARE) between the parabolic fit and the true bands in an energy range of 25 meV. The main computational steps carried out by this Recipe involve three subsequent $k$-point grid refinements; specifically: 
\begin{itemize}
    \item Perform a self-consistent ground state calculation on a uniform $k$-point grid (by calling the \texttt{calculate} Instruction of the Recipe \texttt{asr.gs}) to obtain a converged electron density as well as Kohn--Sham band energies.
    \item Locate the preliminary positions of the VBM and CBM and calculate band energies on a higher-density $k$-point grid around the VBM and CBM to locate the VBM and CBM positions with higher accuracy.
    \item Define final high-density $k$-point grids in the vicinity of the VBM and CBM points, and calculate band energies.
    \item Locate VBM and CBM and fit bands by second-order polynomial using band energies in an energy range of 1 meV from the band extremum. 
    \item Calculate band structures for the web panel and evaluate the ``parabolicity parameter''.
\end{itemize}
It should be noted that even though effective mass calculations appear to be a simple task, it is surprisingly tricky to design a scheme that performs efficiently, robustly, and accurately across all types of band structures including flat bands, highly dispersive bands, highly anisotropic bands, and bands exhibiting complex spin--orbit effects like Rashba splittings.

\subsubsection{General comments}
In contrast to the ``Summary'' panel, which has been customized for the C2DB project (that is, the web panel sections of the relevant Recipes have been appropriately adjusted), the ``Electronic band structure'' and ``Effective masses'' panels are the default web panels produced by the \texttt{asr.bandstructure} and \texttt{asr.emasses} Recipes, respectively. 

The examples given here concern two-dimensional (2D) materials. However, the Recipes \texttt{asr.bandstructure} and \texttt{asr.emasses} (like all other Recipes of the current ASR library) apply also to 1D and 3D materials, as well as 0D where it is meaningful. As mentioned in Section \ref{sec:guideline}, this kind of generality should always be strived for when designing Recipes. Achieving this may be straightforward or more involved depending on the Recipe. The Recipe for the stiffness tensor represents an easy case, where the dimensionality merely dictates the number of axes along which the material must be strained. The Recipe for the band structure is more involved in this regard, as the determination of the band path requires separate treatments in 2D and 3D as does the determination of the spin projection axis (in 2D the out-of-plane direction is a natural choice while in 3D the magnetic easy axis is more appropriate).

\section{User interfaces}\label{sec:interface}
The ASR can be used via four different interfaces, c.f.\ Fig.~\ref{fig:overview}: A command line interface (CLI), a Python interface, a task scheduling front-end, and an app-based interface. Below we describe each interface in more detail. 

\subsection{The CLI} 
The CLI provides convenient commands for easy interaction with ASR via the \texttt{cache} and \texttt{run} subcommands. The \texttt{cache} subcommand allows inspection of the \texttt{Records} stored in the cache, in particular their \texttt{Result} data. For example, \texttt{\$ asr cache ls name=asr.gs} will list all \texttt{Records} produced by the ``ground state'' Recipe. The \texttt{run} subcommand can be used to execute Recipes directly from the command line.  For example, \texttt{\$ asr run asr.gs} will run the \emph{ground state} Recipe.

\subsection{Python interface} 
The Python scripting interface allows inspection of \texttt{Records} and execution of Recipes directly from Python. This makes it possible to implement more complex logic and integrate directly with ASE and any other tools in the user's Python toolkit.

\subsection{MyQueue interface}\label{sec:myqueue} 
For high-throughput computations, ASR can be used in combination with a workflow manager that can handle the interaction with the scheduler of the supercomputer, such as Fireworks\cite{jain2015fireworks} or MyQueue\cite{mortensen2020myqueue}. The latter is a personal, decentralized, and lightweight front-end for schedulers (currently supporting SLURM, PBS, and LSF), which has been co-designed with ASR. MyQueue has a command line interface, which allows for submission of thousands of jobs in one command and provides easy-to-use tools for generating an overview of the status of jobs (`done', `queued', `failed' etc.). It also has a Python interface that can be used to define workflows. A Python script defines a dependency tree of \emph{tasks} that MyQueue will submit without user involvement. The dependencies take the form: ``if task X is done then submit task Y''. MyQueue works directly with folders and files, which makes it transparent and easy to use. Together ASR and MyQueue provide a powerful and extremely flexible toolkit for high-throughput materials computations.

Individual Instructions of the Recipes may be defined as separate MyQueue tasks, such that computational resources can be specifically dedicated each Instruction ensuring a flexible and efficient execution of any workflow. It is, however, not a requirement to specify resources on a per Instruction basis, in which case the resources specified for the main Instruction will apply to all Instructions of the Recipe. 

\subsection{App interface}
The App interface is a web-based read-only interface that allows the user to present and inspect the data stored in an ASE database on a local or public network. Distributing the data on a local network is convenient for larger projects and/or projects involving several users, as it allows for easy sharing and monitoring of the data as the project evolves. Once a project is finalized, the App may be used as a platform to present the data to the world via web pages. The data presentation used by the App is defined in the \texttt{Result} object of the Recipes.  

\section{Data maintenance}\label{sec:data_migration}

It sometimes happens that a Recipe, or one of its Instructions, has to be updated, e.g. because a bug has been detected or it has been found appropriate to store additional metadata. Such updates may imply that previously generated \obj{Records} are no longer consistent with the current implementation of the Recipe. Depending on the nature of the change made to the Recipe, it may be possible to update the \texttt{Record} objects without rerunning the Recipe (data migration) or it may be necessary to rerun the entire Recipe or some of its Instructions (data regeneration).

To support the migration of data, ASR implements a simple versioning system for Instructions. An Instruction is associated with a integer version number which is stored in the \texttt{Record} and identifies the version of the Instruction at the time of creation. When an Instruction is changed, its version number may be increased by the developer. Since the caching layer matches the current Instruction version number against \texttt{Records} in the cache (see Section \ref{sec:record}), older \texttt{Records} would no longer yield cache hits and are then said to be invalidated.

\begin{figure*}[!h]
    \centering
    \includegraphics[width=1.0\textwidth]{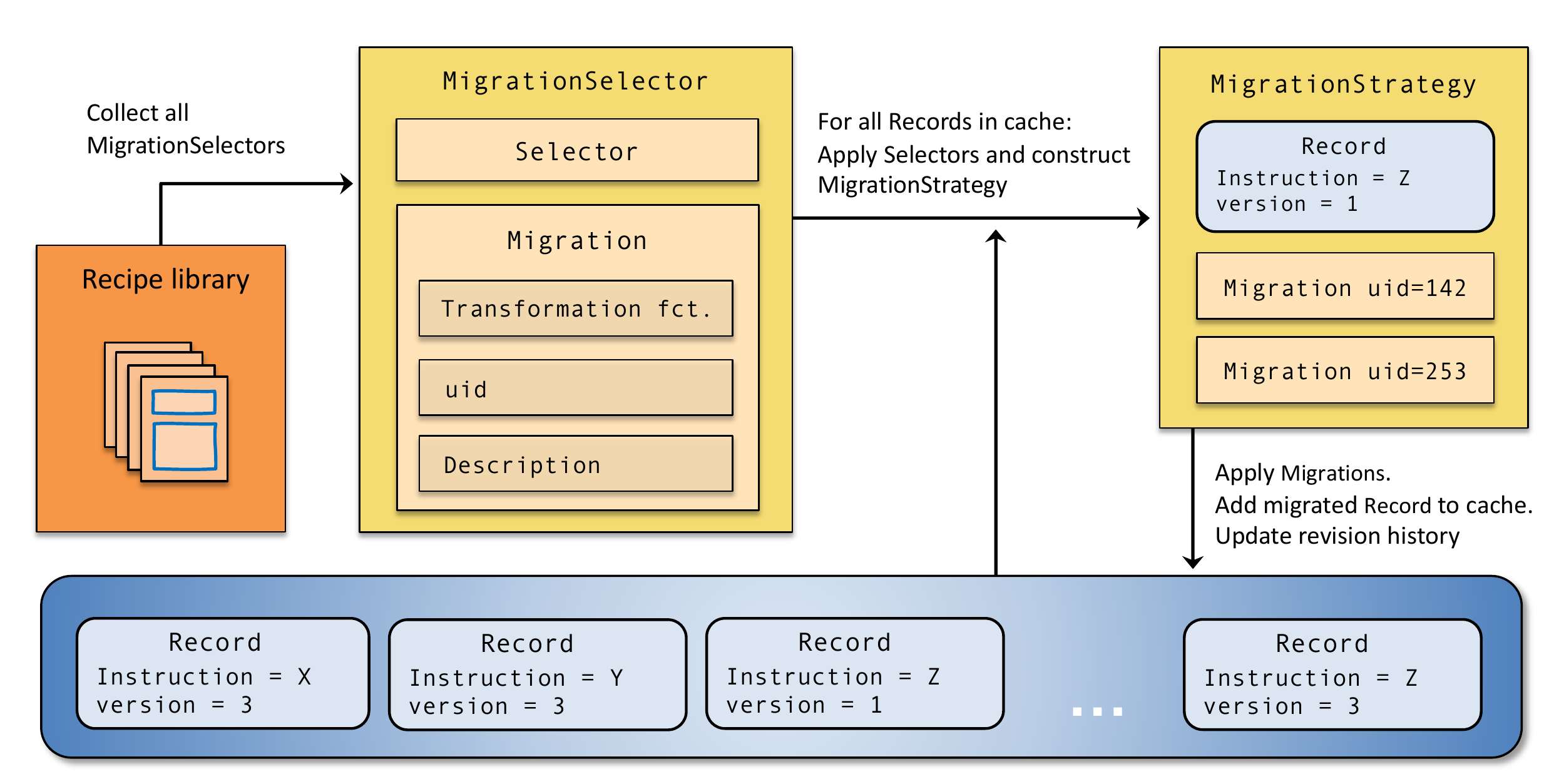}
    \caption{To support data maintenance, ASR provides migration tools for bringing \texttt{Records} up-to-date with the latest version of the Instructions that produced them thereby avoiding recalculations whenever possible. The ASR migration procedure consists of the main steps: (1) Collect all \texttt{MigrationSelectors} from all available \texttt{Instructions}. (2) Select the migratable \texttt{Records} of the cache. (3) Determine a migration strategy (an ordered list of \texttt{Migrations}) for each migratable \texttt{Record}. (4) Apply transformation functions to migrate the \texttt{Records} and add them to the cache. (5) Update the revision history by a \texttt{Revision} object that documents the effect of the migration.}
    \label{fig:migration}
\end{figure*}

To facilitate the migration of invalidated \texttt{Records}, it is possible to specify \texttt{Migrations} that can be associated with an Instruction and thereby provide a way to bring old \texttt{Records} up to date. In practice, a \obj{Migration} bundles a \texttt{Record} transformation function, a unique migration ID and a human readable description of the effect of the migration, see Fig.~\ref{fig:migration}. In general, a \texttt{Record} transformation function induces a change to a \texttt{Record}. For example, this could be to convert a \texttt{Record} of version $n$ to a later version $n+1$ without rerunning the Instruction, but in general the effect of the transformation could be anything. Use of transformation functions is typically possible when the update involves changes to metadata and/or data restructuring while the actual result of the Instruction is unchanged.

When a \texttt{Migration} is applied to a \texttt{Record}, a \texttt{Revision} object is produced. A \texttt{Revision} contains a randomly generated UID, the UID of the applied \texttt{Migration}, an explanatory description of the changes made to the \texttt{Record}, and an automatically generated list of the \texttt{Record} entries that were changed, added or deleted. The auto-generated list of changes is constructed by comparing the \texttt{Record} returned by the transformation function to the input \texttt{Record}. 

Upon migration of a \texttt{Record}, a revision history is updated by the latest \texttt{Revision} and stored in the migrated \texttt{Record}. The revision history can be inspected by users to learn which revisions, if any, have previously been applied to a given \texttt{Record}.

A \texttt{Selector} is used to identify the \texttt{Records} to be migrated, e.g.\ based on the Instruction name and version number. The \texttt{Selector} is bundled together with a \texttt{Migration} into a \texttt{MigrationSelector}, which can determine whether a particular \texttt{Record} matches the selection criteria of the \texttt{Selector}. To migrate a \texttt{Record}, ASR searches through all Recipes to collect their \texttt{MigrationSelectors} (if they have any) and apply them to the \texttt{Record} to find a ``migration strategy'', i.e., which \texttt{Migrations} to be applied and in which order. The migration strategy is then encoded in a \obj{MigrationStrategy}, which couples a particular \texttt{Record} to an ordered list of \texttt{Migrations}. The particular \obj{MigrationStrategy} can then be applied to the Cache to execute the migration of the associated \texttt{Record}.

ASR provides a simple CLI, via \texttt{asr cache migrate}, to analyse existing \texttt{Records} in the cache, and identify migratable \texttt{Records}.

Whenever the \texttt{asr} version used in a given project is upgraded, a project participant should identify migratable \texttt{Records}, migrate them and then rerun the project workflow. Up-to-date \texttt{Records} will then be taken directly from the cache, whereas the Instructions with invalidated \texttt{Records} and no associated \texttt{Migration}, i.e.\ \texttt{Records} that the developer cannot migrate directly to the newest version, will be rerun. 

In order to minimize the computational cost of bringing data up-to-date with ASR, developers are strongly advised to supply \texttt{Migrations} with their Recipe updates whenever possible. 

To provide the best conditions for the long term deployment of ASR-generated data, the \texttt{asr} version of important projects should be upgraded regularly and the project workflow rerun. Obviously, this action may induce changes in the data. Whether this is acceptable or not is ultimately a strategic decision. However, for dynamic data projects, a regular version upgrade not only ensures that the data is of the highest quality, it also makes it easier for other parties to deploy the data because existing results (\texttt{Records}) can be reused directly with the newest version of ASR without having to rerun Recipes to bring the data up-to-date.

\section{Data provenance}\label{sec:data_provenance}
Simply stated, data provenance is the documentation of the circumstances under which a piece of data came into existence. This includes how the data originally was constructed, how the data has changed over time (also known as data-lineage) and a documentation of relevant system specifications such as architecture, operating system, important system packages, executables etc. If data provenance is handled perfectly, then data will in principle be reproducible, i.e.\ given access to exactly the same systems and software, any piece of data can be reproduced. In a scientific context, where reproducibility is key, data provenance is naturally very important.

In ASR, the basic unit of data is the \obj{Record} object, which connects the result of an Instruction with various pieces of contextual metadata, see Section \ref{sec:record}. Taken together, the metadata tell the story of how the original \obj{Record} came into existence (Instruction name/version and input arguments), which other \texttt{Records} were implicitly used for the construction of this \obj{Record} (Dependencies), what external package versions were used, and how the \texttt{Record} has transformed over time (Revision history). For simplicity, since it would be outside the scope of ASR, system information is not stored with the \texttt{Record}, which, in our experience, is not practically relevant for the purposes of ASR. As such, we characterize ASR as practically, but not perfectly, data provenant.

\section{Documentation}\label{sec:documentation}
ASR itself is documented on Read the Docs. The data is documented through the \texttt{Record} and \texttt{Result} objects, see previous Section on data provenance.

 \section{Technical specifications}\label{sec:technical specifications}
Some technical specifications are listed in Table~\ref{tab:specs}.
ASR can be installed via pip using the command \texttt{pip install asr}.
\begin{table}[H]
    \centering
    \begin{tabular}{ll}\toprule
         Source code &  \url{https://gitlab.com/asr-dev/asr}\\
         Releases & \url{https://pypi.org/project/asr/}\\
         License & GNU GPLv3 or newer (free software)\\
         Documentation & \url{https://asr.readthedocs.io/en/latest/}\\
         \toprule
    \end{tabular}
    \caption{Technical specifications}
    \label{tab:specs}
\end{table}

ASR requires or is normally used with the following software:
\begin{itemize}
\item Python libraries: ASE, numpy, matplotlib, plotly, flask, click
\item Computational and workflow software: GPAW or other ASE codes, MyQueue (SLURM/PBS/LSF)
\item Optional extras: spglib, phonopy, and pymatgen (for Recipes); jinja, mysql or other ASE database backends
\end{itemize}
For community support see \url{https://asr.readthedocs.io/en/latest/src/contact.html}.

\section{Summary and outlook}\label{sec:outlook}
This article has introduced The Atomic Simulation Recipes (ASR) as an open source Python framework for developing materials simulation workflows and managing the data they produce. 

To facilitate the transition to a paradigm of data-intensive science, ASR was designed to support the development of materials simulation workflows that operate in accordance with the FAIR data principles, by providing tools and concepts that
are general enough that they do not restrict the user whilst being concrete enough to make a real difference. The ASR achieves this through the notion of a Recipe: a general Python script that performs a well defined simulation task and is wrapped in a caching layer that logs all relevant metadata without involving the user. This construction places essentially no restrictions on the developer's freedom to design and control the workflow, but resolves the critical and complex issue of keeping track of the data provenance. We stress that the core of ASR, i.e.\ the Recipe concept and the caching system, is fully simulation code independent. In particular, it is not tied to materials simulations and could potentially be useful in other areas of computational science. 

Beyond the built-in data documentation, there are many benefits of using standardized, well tested, and well documented Recipes. For example, it saves time and promotes a more sustainable scripting culture by reducing the need for individual researchers to write and maintain their own personal scripts (which can be hard for other to read and are often lost when the developer leaves the group). Furthermore, it reduces the risk of human errors and lowers the barrier for researchers to undertake simulation tasks with which they have little prior experience.  

The fact that Recipes are independent units with own data provenance control implies that they can be freely combined to create advanced workflows using Python scripting for maximal flexibility. Such workflows can be executed on supercomputers using a workflow management software that supports a Python interface. To this end, we have developed the MyQueue\cite{mortensen2020myqueue} task manager that works as a front-end to the most common schedulers (currently SLURM, PBS, and LSF). While MyQueue will resubmit jobs that have timed out or crashed due to lack of memory, code-related failures must be handled manually. In the future, ASR should integrate more closely with MyQueue to permit that errors from the simulation codes are automatically analysed and reacted upon. Along the same lines, an automated estimation of the HPC resources (time/memory/nodes) required by individual tasks could limit the number of failed jobs and improve the utilization of resources.    

The current Recipe library already covers a wide range of materials simulation tasks and more are continuously being added. Of special importance are Recipes for advanced beyond-DFT calculations where the benefits in terms of a lowered user barrier, improved data quality, and increased utilization of computing resources, are particularly large. 
The Recipe concept should also be advantageous for implementation of machine learning methods that could integrate with ASR databases and ``standard'' Recipes to make for more intelligent and computationally efficient workflows.   

The ASR makes extensive use of the Atomic Simulation Environment (ASE) as a toolkit to process atomistic calculations. In particular, ASE is used as a front-end for ASR to communicate with
external simulation codes.
This has the clear advantage that ASR can become decoupled from the simulation codes. This decoupling is currently not in place, and the majority of the existing Recipe implementations contain code parts that are specific to the GPAW electronic structure code. To make ASR fully simulation code-independent, the ASE Calculator interfaces must be further generalized. This includes extensions of the interfaces to access outputs of
calculations as well as a systematic mechanism
to control multi-step tasks.
The adaptation of this interface to multiple codes will eventually
require a community effort that we hope many code developers will take
part in.
Until then, Recipes must to some extent be code specific.

\section{Acknowledgments}
We acknowledge funding from the European Research Council (ERC)  under  the  European  Union’s  Horizon  2020  research and innovation program Grant No. 773122 (LIMA) and Grant agreement No. 951786 (NOMAD CoE).

\bibliographystyle{iopart-num}
\bibliography{references}

\end{document}